\documentclass[12pt]{article}

\usepackage[totalwidth=460truept,totalheight=600truept]{geometry}
\usepackage{latexsym,graphicx,amsfonts,amssymb,amsmath,xcolor,changepage}
\usepackage[hypertexnames=false,hidelinks]{hyperref}

\global\arraycolsep=1truept

\null

\begin{document}

\medskip 

\begin{center}
{\Huge \textbf{Propagators and Widths of}}

\vskip.6truecm

{\Huge \textbf{Physical and Purely Virtual Particles}}

\vskip.55truecm

{\Huge \textbf{in a Finite Interval of Time}}

\vskip.5truecm
\end{center}

\vskip.5truecm

\begin{center}
\textsl{Damiano Anselmi}

\vskip.1truecm

{\small \textit{Dipartimento di Fisica \textquotedblleft
E.Fermi\textquotedblright , Universit\`{a} di Pisa, Largo B.Pontecorvo 3,
56127 Pisa, Italy}}

{\small \textit{INFN, Sezione di Pisa, Largo B. Pontecorvo 3, 56127 Pisa,
Italy}}

{\small damiano.anselmi@unipi.it}

\vskip .5truecm

\textbf{Abstract}
\end{center}

We study the free and dressed propagators of physical and purely virtual
particles in a finite interval of time $\tau $ and on a compact space
manifold $\Omega $, using coherent states. In the free-field limit, the
propagators are described by the entire function $(e^{z}-1-z)/z^{2}$, whose
shape on the real axis is similar to the one of a Breit-Wigner function,
with an effective width around $1/\tau $. The real part is positive, in
agreement with unitarity, and remains so after including the radiative
corrections, which shift the function into the physical half plane. We
investigate the effects of the restriction to finite $\tau $ on the problem
of unstable particles vs resonances, and show that the muon observation
emerges from the right physical process, differently from what happens at $%
\tau =\infty $. We also study the case of purely virtual particles, and show
that, if $\tau $ is small enough, there exists a situation where the
geometric series of the self-energies is always convergent. The plots of the
dressed propagators show testable differences: while physical particles are
characterized by the usual, single peak, purely virtual particles are
characterized by twin peaks.

\vfill\eject

\section{Introduction}

\label{intro}\setcounter{equation}{0}

Widths are key quantities in quantum field theory, and a link between
perturbative and nonperturbative quantum field theory. A perturbatively
stable particle may decay after the resummation of its self-energies into
the so-called dressed propagator. Yet, the resummation, which is normally
considered a straightforward operation, has unexpected features, when it
comes to explain the observation of long-lived unstable particles, like the
muon \cite{fSelfK}.

The $S$ matrix amplitudes allow us to study scattering processes between
asymptotic states, which are separated by an infinite amount of time. In
this scenario, a long-lived unstable particle always has enough time to
decay, before being actually observed. Although it is possible to make room
for the muon observation in a rough and ready way within the usual
frameworks, too many important details are missed along the way by doing so.
It is much better to study the problem where it belongs, which is quantum
field theory in a finite interval of time.

It is possible to formulate quantum field theory in a finite time interval $%
\tau $, and on a compact space manifold $\Omega $, by moving most details
about such restrictions away from the internal sectors of the diagrams into
external sources \cite{MQ}. Then the diagrams are the same as usual, apart
from the discretization of the loop momenta, and the presence of sources
attached to the vertices. Most known properties of the usual $S$ matrix
amplitudes generalize straightforwardly, and allow us to study the
systematics of renormalization and unitarity \cite{MQ}. The formulation is
well-suited to be generalized so as to include purely virtual particles,
i.e., particles that do not exist on the mass shell at any order of the
perturbative expansion. At $\tau =\infty $, $\Omega =\mathbb{R}^{3}$, they
are introduced by removing the on-shell contributions of a physical particle 
$\chi _{\text{ph}}$ (or a ghost $\chi _{\text{gh}}$, which is a particle
with the wrong sign in front of its kinetic term) from the internal parts of
the diagrams \cite{diagrammarMio}, and restricting to the diagrams that do
not contain $\chi _{\text{ph}}$, $\chi _{\text{gh}}$ on the external legs.
At finite $\tau $ and on a compact $\Omega $, they are introduced by
removing the same on-shell parts from the core diagrams, and choosing
trivial initial and final conditions for $\chi _{\text{ph}}$, $\chi _{\text{%
gh}}$ \cite{MQ}. The evolution operator of the resulting theory is unitary,
provided all the ghosts are rendered purely virtual.

In this paper, we study the propagators of physical and purely virtual
particles in a finite interval of time $\tau $, and on a compact space
manifold $\Omega $. In the free-field limit, the typical pole $1/z$ of the
usual propagator at $\tau =\infty $ is replaced by an entire function, which
is $f(z)=(e^{z}-1-z)/z^{2}$. Although $f(z)$ is very different from $1/z$
(and from a Breit-Wigner function) in most of the complex plane, its shape
on the real axis $z=ix$, $x\in \mathbb{R}$, does remind the one of a
Breit-Wigner function, with an effective width equal to $16/(3\tau )$. When
we include the radiative corrections, the function $f(z)$ is shifted into
the physical half plane, where the real part of the propagator remains
positive, consistently with unitarity. The width is enlarged by an amount
equal to $\Gamma $ (the usual width at $\tau =\infty $).

The muon observation emerges rather naturally from the right physical
process: there is no need to confuse the observation of an unstable particle
with the observation of its decay products, as one normally does to adjust
the matter at $\tau =\infty $.

In the case of purely virtual particles, we show that, for $\tau $ small
enough, there is an arrangement where the geometric series of the
self-energies is always convergent. In that situation, we can resum the
radiative corrections rigorously to the very end, and obtain the dressed
propagator. Comparing the plot of its real part with the one of physical
particles, testable differences emerge: while the physical particles are
characterized by the common, single peak, purely virtual particles are
characterized by two twin peaks.

The results confirm the ones of ref. \cite{fSelfK}, where they were derived
by arguing, on general grounds, what the main effects of the restriction to
finite $\tau $ were going to be.

Both physical particles and ghosts can be rendered purely virtual. At the
same time, purely virtual particles are not Lee-Wick ghosts \cite{leewick}%
\footnote{%
For Lee-Wick ghosts in quantum gravity, see \cite{tomboulis}}, as shown in 
\cite{LWfakeons}. In particular, they do not need to have nonvanishing
widths, and decay. And even if they have a nonvanishing width $\Gamma _{%
\text{f}}$, its meaning is not the reciprocal of a lifetime, nor the actual
width of a peak. In the case studied here, where the resummation of the
dressed propagator can be done rigorously to the very end, $\Gamma _{\text{f}%
}$ is a measure of the height of the twin peaks, while their distance is
universally fixed to $2\pi $ (in suitable units). In every other case, the
\textquotedblleft peak region\textquotedblright\ of a purely virtual
particle is nonperturbative. Certain arguments suggest that $\Gamma _{\text{f%
}}$ may measure a \textquotedblleft peak uncertainty\textquotedblright\ $%
\Delta E>\Gamma _{\text{f}}/2$, telling us that, when we approach the peak
region too close in energy, identical experiments may give different results 
\cite{fSelfK}. 

At the phenomenological level, purely virtual particles may have other
interesting applications, because they evade many constraints that are
typical of normal particles (see \cite{Tallinn1,Tallinn2,PivaMelis} and
references therein).

The paper is organized as follows. In section \ref{freepropa} we study the
free propagator at finite $\tau $. In section \ref{dressedpropa} we resum
the self-energies into the dressed propagator. In section \ref{PV} we study
the free and dressed propagators of purely virtual particles. In section \ref%
{muon} we investigate the problem of unstable particles. Section \ref%
{conclusions} contains the conclusions. We work on bosonic fields, since the
generalization to fermions and gauge fields does not present problems.

\section{Free propagator in a finite interval of time}

\label{freepropa}\setcounter{equation}{0}

In this section, we study the free propagator in a finite interval of time $%
\tau $. For most purposes of this paper, we can Fourier transform the space
coordinates, understand the integrals on the loop momenta, and concentrate
on time and energy. This means that we can basically work with quantum
mechanics, where the coordinates $q(t)$ stand for fields $\phi (t,\mathbf{x}%
) $. We assume that the Lagrangian has the form 
\begin{equation}
L_{\lambda }(q,\dot{q})=\frac{1}{2}\left( \dot{q}^{2}-\omega
^{2}q^{2}\right) -V_{\lambda }(q,\dot{q}),  \label{ll}
\end{equation}%
where $V_{\lambda }(q)$ is proportional to some coupling $\lambda $. If the
space manifold $\Omega $ is compact, the frequencies $\omega $ are
restricted to a discrete set $\omega _{\mathbf{n}}$, for some label $\mathbf{%
n}$. This affects the propagator only in a minor way. Effects like these
will be understood, from now on, so the formulas we write look practically
the same as on $\Omega =\mathbb{R}^{3}$.

We use coherent \textquotedblleft states\textquotedblright\ \cite{cohe} (so
we call them, although we work in the functional-integral approach)%
\begin{equation}
z=\frac{1}{2}\left( q+i\frac{p}{\omega }\right) ,\qquad \bar{z}=\frac{1}{2}%
\left( q-i\frac{p}{\omega }\right) ,  \label{chvcoh}
\end{equation}%
where $p=\partial L_{\lambda }/\partial \dot{q}$ is the momentum\footnote{%
We use the notation of \cite{MQ}, where details about the switch to coherent
states can be found.}. So doing, we double the number of coordinates, or
fields, lower the number of time derivatives from two to one, and treat the
poles of the propagator%
\begin{equation}
\frac{i}{k^{2}-m^{2}+i\epsilon }=\frac{i}{2\omega }\left( \frac{1}{e-\omega
+i\epsilon }-\frac{1}{e+\omega -i\epsilon }\right)  \label{pinf}
\end{equation}%
separately\footnote{%
A redefinition on $\epsilon $ is understood between the left- and right-hand
sides of (\ref{pinf}).}, where $k^{\mu }=(e,\mathbf{k})$ is the
four-momentum and $\omega =\sqrt{\mathbf{k}^{2}+m^{2}}$ denotes the
frequency.

The first pole gives the propagator 
\begin{equation}
G^{+}(t,t^{\prime })=\langle z(t)\hspace{0.01in}\bar{z}(t^{\prime })\rangle
_{0}=\theta (t-t^{\prime })\frac{\mathrm{e}^{-i\omega (t-t^{\prime })}}{%
2\omega },  \label{ppinf}
\end{equation}%
while the other pole gives $G^{-}(t,t^{\prime })=\langle \bar{z}(t)\hspace{%
0.01in}z(t^{\prime })\rangle _{0}=G^{+}(t^{\prime },t)$. Moreover, $\langle
z(t)\hspace{0.01in}z(t^{\prime })\rangle _{0}=\langle \bar{z}(t)\hspace{%
0.01in}\bar{z}(t^{\prime })\rangle _{0}=0$. The sum%
\begin{equation}
G(t,t^{\prime })=G^{+}(t,t^{\prime })+G^{-}(t,t^{\prime })=\frac{\mathrm{e}%
^{-i\omega |t-t^{\prime }|}}{2\omega }  \label{fey}
\end{equation}%
is indeed the Fourier transform of the Feynman propagator (\ref{pinf}).

When $\tau =\infty $, the propagators are (\ref{pinf}) and (\ref{ppinf}) for
all real values of $t$ and $t^{\prime }$. When $\tau $ is finite, the
propagators are unaffected, in the coherent-state approach, apart from the
restrictions of $t$ and $t^{\prime }$ to the interval $(t_{\text{i}},t_{%
\text{f}})$. To make this restriction explicit, we multiply both sides of $%
G^{\pm }(t,t^{\prime })$ and $G(t,t^{\prime })$ by projectors $\Pi _{\tau
}(t)\equiv \theta (t_{\text{f}}-t)\theta (t-t_{\text{i}})$ and $\Pi _{\tau
}(t^{\prime })$. The projected propagators are then 
\begin{equation}
G_{\tau }(t,t^{\prime })=\Pi _{\tau }(t)G(t,t^{\prime })\Pi _{\tau
}(t^{\prime }),\qquad G_{\tau }^{\pm }(t,t^{\prime })=\Pi _{\tau }(t)G^{\pm
}(t,t^{\prime })\Pi _{\tau }(t^{\prime }).  \label{proje}
\end{equation}%
For simplicity, we take $t_{\text{f}}=\tau /2$, $t_{\text{i}}=-\tau /2$.

It is interesting to study the Fourier transforms of (\ref{proje}), which
can be calculated by assuming that $\omega $ has a small, negative imaginary
part. We start from%
\begin{equation}
\tilde{G}_{\tau }^{+}(e,e^{\prime })=\int_{-\infty }^{+\infty }\mathrm{d}%
t\int_{-\infty }^{+\infty }\mathrm{d}t^{\prime }\hspace{0.01in}G_{\tau
}^{+}(t,t^{\prime })\mathrm{e}^{i(et+e^{\prime }t^{\prime })}.  \label{fu}
\end{equation}%
Due to the lack of invariance under time translations, the result does not
factorize the usual energy-conservation delta function $(2\pi )\delta
(e+e^{\prime })$. Instead, we can factorize a%
\begin{equation}
\frac{2\sin \left( \frac{e+e^{\prime }}{2}\tau \right) }{e+e^{\prime }},
\label{ff}
\end{equation}%
which is the Fourier transform of $\Pi _{\tau }(t)$ with energy $e+e^{\prime
}$. Furthermore, we assume that $\tau $ is large enough, so that we can
restrict the coefficient of (\ref{ff}) in $\tilde{G}_{\tau }^{+}(e,e^{\prime
})$ to $e+e^{\prime }=0$. Factorizing a $2\omega /\tau $ for convenience, we
approximate $\tilde{G}_{\tau }^{+}(e,e^{\prime })$ to%
\begin{equation}
\tilde{G}_{\tau }^{+}(e,e^{\prime })\simeq \frac{2\sin \left( \frac{%
e+e^{\prime }}{2}\tau \right) }{e+e^{\prime }}\frac{\tau }{2\omega }f(z),
\label{Gtauplus}
\end{equation}%
where $z=i(e-\omega )\tau $. We find 
\begin{equation*}
f(z)=2\lim_{e^{\prime }\rightarrow -e}\frac{(e+e^{\prime })\omega \tilde{G}%
_{\tau }^{+}(e,e^{\prime })}{\tau \sin \left( \frac{e+e^{\prime }}{2}\tau
\right) }=\frac{2\omega }{\tau ^{2}}\lim_{e^{\prime }\rightarrow -e}\tilde{G}%
_{\tau }^{+}(e,e^{\prime })=\frac{\mathrm{e}^{z}-1-z}{z^{2}}.
\end{equation*}%
Interestingly enough, $f(z)$ is an entire function: the propagator at finite 
$\tau $ has no pole, and no other type of singularity.

Writing $z=ix$, it is useful to single out the real and imaginary parts:%
\begin{equation*}
f(z)=\frac{1-\cos (x)}{x^{2}}+i\frac{x-\sin (x)}{x^{2}}.
\end{equation*}%
To verify that the limit $\tau \rightarrow \infty $ gives the usual result,
we first rescale $\tau $ by a factor $\lambda $ and then let $\lambda $ tend
to infinity by means of the identities%
\begin{equation}
\lim_{\lambda \rightarrow \infty }\frac{\sin \left( \lambda x\right) }{x}%
=\pi \delta (x),\qquad \lim_{\lambda \rightarrow \infty }\frac{1-\cos
(\lambda x)}{\lambda x^{2}}=\pi \delta (x),\qquad \lim_{\lambda \rightarrow
\infty }\frac{\lambda x-\sin (\lambda x)}{\lambda x^{2}}=\mathcal{P}\frac{1}{%
x}  \label{distro}
\end{equation}%
(which can be easily proved by studying them on test functions), where $%
\mathcal{P}$ denotes the Cauchy principal value. Thus,%
\begin{equation*}
\lim_{\tau \rightarrow \infty }\tilde{G}_{\tau }^{+}(e,e^{\prime })=(2\pi
)\delta (e+e^{\prime })\frac{1}{2\omega }\frac{i}{e-\omega +i\epsilon }=%
\tilde{G}^{+}(e,e^{\prime }).
\end{equation*}%
Summing it to $\tilde{G}_{\tau }^{-}(e,e^{\prime })=\tilde{G}_{\tau
}^{+}(e^{\prime },e)$, we go back to the Feynman propagator at $\tau =\infty 
$:%
\begin{equation}
\tilde{G}(e,e^{\prime })=\tilde{G}^{+}(e,e^{\prime })+\tilde{G}%
^{-}(e,e^{\prime })=(2\pi )\delta (e+e^{\prime })\frac{i}{%
k^{2}-m^{2}+i\epsilon }.  \label{usualF}
\end{equation}

Finally, the Fourier transform of the total propagator~$G_{\tau
}(t,t^{\prime })$ at finite $\tau $ is%
\begin{equation}
\tilde{G}_{\tau }(e,e^{\prime })\simeq \frac{2\tau ^{2}\sin \left( \frac{%
e+e^{\prime }}{2}\tau \right) }{e+e^{\prime }}h(x_{+},x_{-}),
\label{undrepro}
\end{equation}%
where%
\begin{equation*}
h(x_{+},x_{-})\equiv \frac{f(ix_{+})+f(-ix_{-})}{x_{-}-x_{+}},\qquad x_{\pm
}\equiv (e\mp \omega )\tau .
\end{equation*}

\bigskip

\begin{figure}[t]
\begin{center}
\includegraphics[width=5truecm]{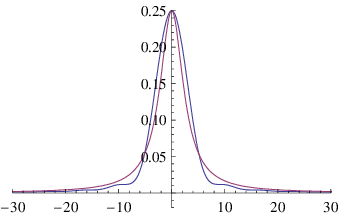}\quad %
\includegraphics[width=5truecm]{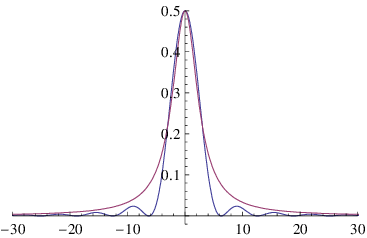}\quad %
\includegraphics[width=5truecm]{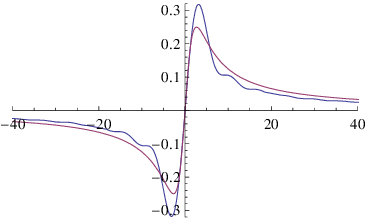}
\end{center}
\par
\vskip-.5truecm
\caption{Comparison between $f(z)$ (in blue) and $f_{\text{BW}}(z)$ (in
red): square modulus (left), real part (middle) and imaginary part (right)}
\label{BW}
\end{figure}

We see that the propagator at finite $\tau $ is encoded into the key
function $f(z)$. It is convenient to compare it to a \textquotedblleft
twin\textquotedblright\ Breit-Wigner (BW) function $f_{\text{BW}}(z)$,
determined so that $f(z)$ and $f_{\text{BW}}(z)$ have the same values at $%
z=0 $ and the same $L^{2}(\mathbb{R})$ norms on the real axis (by which we
mean for $z=ix$, $x\in \mathbb{R}$). We find%
\begin{equation}
f_{\text{BW}}(z)=\frac{4}{8-3z},\qquad \lim_{z\rightarrow 0}f_{\text{BW}%
}(z)=\lim_{z\rightarrow 0}f(z),\qquad \int_{-\infty }^{+\infty }|f_{\text{BW}%
}(ix)|^{2}\mathrm{d}x=\int_{-\infty }^{+\infty }|f(ix)|^{2}\mathrm{d}x.
\label{BWa}
\end{equation}%
The width $\Gamma _{\text{eff}}$ of the twin\ function $f_{\text{BW}}(z)$ is
a good measure of the effective width of the function $f(z)$ on the real
axis. We find 
\begin{equation}
\Gamma _{\text{eff}}=\frac{16}{3\tau },\qquad f_{\text{BW}}(i(e-\omega )\tau
)=\frac{4i}{3\tau \left( e-\omega +i\frac{\Gamma _{\text{eff}}}{2}\right) }.
\label{effGamma}
\end{equation}%
In fig. \ref{BW} we compare the square moduli, the real parts and the
imaginary parts of $f(z)$ and $f_{\text{BW}}(z)$. We see that their slices
on the real axis are similar, although the functions differ a lot in the
rest of the complex plane.

It is also possible to approximate the total propagator (\ref{undrepro}) by
replacing the function $f$ with the twin BW function $f_{\text{BW}}$. The
approximation is good enough when the distance $x_{-}-x_{+}$ between the two
peaks (the one of the particle and the one of the antiparticle) is large.
When $x_{-}-x_{+}$ decreases, effects due to the superposition between the
two peaks start to become important, although the approximation remains good
qualitatively. 

\begin{figure}[t]
\begin{center}
\includegraphics[width=7truecm]{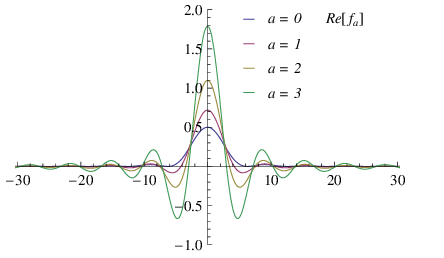}\quad %
\includegraphics[width=7truecm]{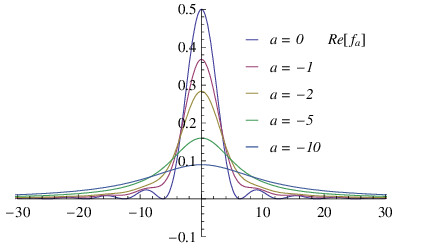}\\[0pt]
\includegraphics[width=7truecm]{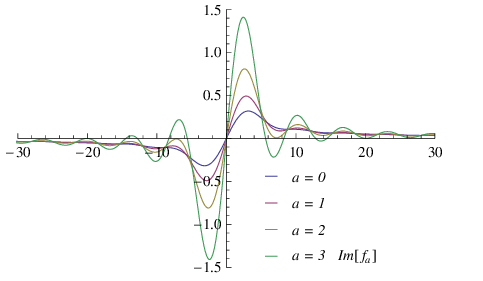}\quad %
\includegraphics[width=7truecm]{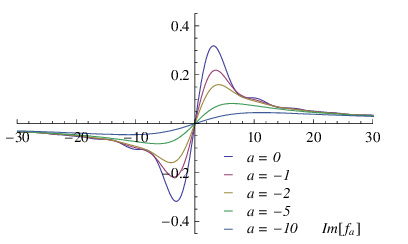}
\end{center}
\par
\vskip-.5truecm
\caption{Real and imaginary parts of the function $f_{a}(x)=f(ix+a)$ for
various values of $a$}
\label{fa}
\end{figure}

Now we describe $f(z)$ for generic complex $z$. We shift $z=ix$ by a real
constant $a$, with $x$ also real, and compare parallel slices $%
f_{a}(ix)\equiv f(ix+a)$. The typical behaviors of the real and imaginary
parts of $f_{a}(ix)$ are shown in figure \ref{fa}, for positive and negative 
$a$. We see that the real part is always positive for $a<0$, but can have
both signs for $a>0$. 

For $a<0$, the function Re$[f_{a}(ix)]$ still looks like the real part of a
BW function, but with a larger width. The physical meaning of this behavior
is explained by the radiative corrections. Specifically, we show that a
negative $a$ originates from the resummation of the self-energy diagrams
into the dressed propagator, and is ultimately proportional to $-\Gamma $,
where $\Gamma $ is the usual particle width.

\section{Dressed propagator}

\label{dressedpropa}\setcounter{equation}{0}

In this section we study the dressed propagator, by resumming the
corrections due to the self-energy diagrams.

Let $\Sigma (t,t^{\prime })$ denote the usual self-energy (at $\tau =\infty $%
) and $\Sigma _{\tau }(t,t^{\prime })$ the one at finite $\tau $. For what
we are going to say, it is sufficient to focus on the one-loop corrections
in the simplest case, where $-i\Sigma (t,t^{\prime })$ is the bubble diagram
in coordinate space (e.g., the product of two propagators between the same,
non coinciding points). Then, $\Sigma _{\tau }(t,t^{\prime })=\Pi _{\tau
}(t)\Sigma (t,t^{\prime })\Pi _{\tau }(t^{\prime })$.

The dressed propagator $G_{\tau \text{d}}(t,t^{\prime })$, obtained from the
mentioned resummation, reads 
\begin{eqnarray}
G_{\tau \text{d}}(t,t^{\prime }) &=&\Pi _{\tau }(t)\hat{G}_{\tau \text{d}%
}(t,t^{\prime })\Pi _{\tau }(t^{\prime }),  \notag \\
\hat{G}_{\tau \text{d}}(t,t^{\prime }) &\equiv &G(t,t^{\prime })+\int 
\mathrm{d}t_{1}\mathrm{d}t_{2}G(t,t_{1})(-i)\Sigma _{\tau
}(t_{1},t_{2})G(t_{2},t^{\prime })  \notag \\
&&\!\!\!\!\!\!\!\!\!\!\!\!{+\int \mathrm{d}t_{1}\mathrm{d}t_{2}\mathrm{d}%
t_{3}\mathrm{d}t_{4}G(t,t_{1})(-i)\Sigma _{\tau
}(t_{1},t_{2})G(t_{2},t_{3})(-i)\Sigma _{\tau
}(t_{3},t_{4})G(t_{4},t^{\prime })+\cdots ,}\qquad  \label{gtd}
\end{eqnarray}%
where $\hat{G}_{\tau \text{d}}(t,t^{\prime })$\ is a sort of unprojected
dressed propagator.

We can work out the resummation in two ways, which are equivalent within the
approximations we are making here.

In the first method we first show that $\Sigma _{\tau }$ can be replaced by $%
\Sigma $ inside $\hat{G}_{\tau \text{d}}(t,t^{\prime })$. This makes $\hat{G}%
_{\tau \text{d}}(t,t^{\prime })$ coincide with the usual dressed propagator $%
G_{\text{d}}(t,t^{\prime })$ at $\tau =\infty $. Then, $G_{\tau \text{d}%
}(t,t^{\prime })$ is the projected version of $G_{\text{d}}(t,t^{\prime })$,
which can be worked out as we did in the previous section.

In Fourier transforms, the usual bubble diagram can be approximated by a
constant around the peak, which encodes the mass renormalization $\Delta
m^{2}$ and the (nonnegative) width $\Gamma $:%
\begin{equation}
\Sigma (t,t^{\prime })\simeq \delta (t-t^{\prime })(\Delta m^{2}-im_{\text{ph%
}}\Gamma ),\qquad \tilde{\Sigma}(e,e^{\prime })\simeq (2\pi )\delta
(e+e^{\prime })(\Delta m^{2}-im_{\text{ph}}\Gamma ),  \label{approxa}
\end{equation}%
where $m_{\text{ph}}^{2}\equiv m^{2}+\Delta m^{2}$. We ignore the radiative
corrections to the normalization factor $Z$ of the propagator, since we can
reinstate $Z$ at a later time. Using the approximation (\ref{approxa}) as
the whole self-energy, the Fourier transform of $\Sigma _{\tau }(t,t^{\prime
})$ is%
\begin{equation}
\tilde{\Sigma}_{\tau }(e,e^{\prime })\simeq \frac{2\sin \left( \frac{%
e+e^{\prime }}{2}\tau \right) }{e+e^{\prime }}(\Delta m^{2}-im_{\text{ph}%
}\Gamma ).  \label{approza}
\end{equation}

As before, we neglect the energy nonconservation at the vertices, by
assuming that $\tau $ is large enough so that we can replace the factor in
front by $(2\pi )\delta (e+e^{\prime })$. We obtain $\tilde{\Sigma}_{\tau
}(e,e^{\prime })\simeq \tilde{\Sigma}(e,e^{\prime })$, which means that the
restriction to finite $\tau $ has negligible effects on $\hat{G}_{\tau \text{%
d}}(t,t^{\prime })$, and we can replace it with $G_{\text{d}}(t,t^{\prime })$%
. Then (\ref{gtd}) gives 
\begin{equation}
G_{\tau \text{d}}(t,t^{\prime })=\Pi _{\tau }(t)G_{\text{d}}(t,t^{\prime
})\Pi _{\tau }(t^{\prime }).  \label{gtd2}
\end{equation}

Resumming the geometric series, the Fourier transform of $G_{\text{d}%
}(t,t^{\prime })$, which is%
\begin{equation}
\tilde{G}_{\text{d}}(e,e^{\prime })=(2\pi )\delta (e+e^{\prime })\frac{i}{%
k^{2}-m_{\text{ph}}^{2}+im_{\text{ph}}\Gamma },  \label{dressedF}
\end{equation}%
is the same as $\tilde{G}(e,e^{\prime })$, formula (\ref{usualF}), with the
replacement $m^{2}\rightarrow m_{\text{ph}}^{2}-im_{\text{ph}}\Gamma $.
Then, by comparing the first formula of (\ref{proje}) with (\ref{gtd2}), and
using (\ref{undrepro}), we conclude that the Fourier transform of $G_{\tau 
\text{d}}(t,t^{\prime })$ is%
\begin{equation}
\tilde{G}_{\tau \text{d}}(e,e^{\prime })\simeq \left. \tilde{G}_{\tau
}(e,e^{\prime })\right\vert _{m^{2}\rightarrow m_{\text{ph}}^{2}-im_{\text{ph%
}}\Gamma }=\frac{2\tau ^{2}\sin \left( \frac{e+e^{\prime }}{2}\tau \right) }{%
e+e^{\prime }}h(x_{+\text{ph}},x_{-\text{ph}}),  \label{tota}
\end{equation}%
where%
\begin{equation*}
\tilde{\omega}_{\text{ph}}=\sqrt{\omega _{\text{ph}}^{2}-im_{\text{ph}%
}\Gamma },\qquad \omega _{\text{ph}}=\sqrt{\mathbf{k}^{2}+m_{\text{ph}}^{2}},%
\text{\qquad }x_{\pm \text{ph}}\equiv (e\mp \tilde{\omega}_{\text{ph}})\tau .
\end{equation*}

We see that we just need to make the replacements%
\begin{equation*}
\pm ix_{\pm }\rightarrow \pm ix_{\pm \text{ph}}=\pm ix_{\pm }+a+ib,\qquad
a+ib=i\tau \left( \omega -\tilde{\omega}_{\text{ph}}\right) ,
\end{equation*}
inside the functions $f$, with $a$, $b\in \mathbb{R}$. While $b$ is a\
simple translation of $x_{\pm }$, the quantity%
\begin{equation}
a=\tau \text{Im\hspace{0.01in}}[\tilde{\omega}_{\text{ph}}]<0  \label{aa}
\end{equation}%
measures the displacement of the plot profile into the physical half plane.
Assuming $\Gamma \ll m_{\text{ph}}$, we have $a\simeq -\tau m_{\text{ph}%
}\Gamma /(2\omega _{\text{ph}})$. Moreover, $a\simeq -\tau \Gamma /2$ in the
static limit.

The total propagator $\tilde{G}_{\tau \text{d}}(e,e^{\prime })$ is described
by the function%
\begin{equation}
g(x,y,a)=\frac{f(i(x-y)+a)+f(-i(x+y)+a)}{2(y+ia)},\qquad y>-a>0,
\label{repro}
\end{equation}%
where $x=(x_{+}+x_{-})/2=\tau e$, $y=(x_{-}-x_{+}-2b)/2=\omega \tau -b=\tau 
\hspace{0.01in}$Re$[\tilde{\omega}_{\text{ph}}]$. The quantity $2y$ is a
measure of the separation between the particle peak and the antiparticle
peak. If we choose $y$ large enough, we can compare the properties of the
two resummation methods more clearly, because we avoid the superposition of
the two peaks. The validity of our results does not depend on this
assumption.

\begin{figure}[t]
\begin{center}
\includegraphics[width=5truecm]{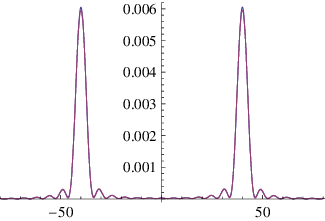}\quad %
\includegraphics[width=5truecm]{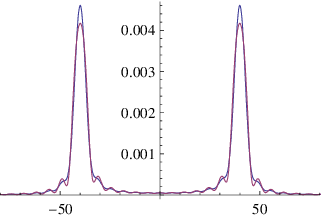}\quad %
\includegraphics[width=5truecm]{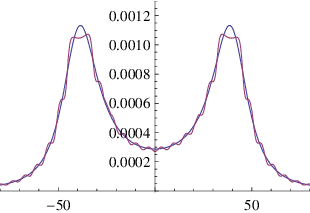}
\end{center}
\par
\vskip-.5truecm
\caption{Plots of Re$[g(x,y,a)]$ (in blue) and Re$[g^{\prime }(x,y,a)]$ (in
red) for $y=40$, $a=-1/10$ (left), $a=-1$ (middle) and $a=-10$ (right).}
\label{2peak}
\end{figure}

The parameter $-a$ measures the extra width due to the radiative
corrections. The inequality $y+a>0$ holds because $\tilde{\omega}_{\text{ph}%
}^{2}$ lies in the fourth quadrant of the complex plane, so Re$[\tilde{\omega%
}_{\text{ph}}]+$ Im\hspace{0.01in}$[\tilde{\omega}_{\text{ph}}]>0$. It is
easy to check that the real part of $g(x,y,a)$ is positive, in agreement
with unitarity (see fig. \ref{2peak}). 

The second way of resumming the self-energies amounts to working directly on
the Fourier transforms, by means of (\ref{undrepro}) and (\ref{approza}).
For simplicity, we assume $\Delta m^{2}=0$ from now on, since the mass
redefinition is not crucial for what we are going to say. We take care of
the energy conservation by approximating the factor $(2/(e+e^{\prime }))\sin
\left( (e+e^{\prime })\tau /2\right) $ to $(2\pi )\delta (e+e^{\prime })$
everywhere in the sum, and switching back to the original factor only in the
final formula. Then we get a straightforward geometric series, which sums to%
\begin{equation}
\frac{(e+e^{\prime })\tilde{G}_{\tau \text{d}}^{\prime }(e,e^{\prime })}{%
2\tau ^{2}\sin \left( \frac{e+e^{\prime }}{2}\tau \right) }\simeq
h(x_{+},x_{-})\sum_{n=0}^{\infty }\left( -m\Gamma \tau
^{2}h(x_{+},x_{-})\right) ^{n}\equiv g^{\prime }(x,y,a).  \label{gp}
\end{equation}%
We find 
\begin{equation}
g^{\prime }(x,y,a)=\frac{f(i(x-\bar{y}))+f(-i(x+\bar{y}))}{2\bar{y}\left(
1+\gamma f(i(x-\bar{y}))+\gamma f(-i(x+\bar{y}))\right) },  \label{gpp}
\end{equation}%
where%
\begin{equation*}
x=\tau e,\qquad \gamma =\frac{\tau m\Gamma }{2\omega }=-\frac{ay}{\bar{y}}%
,\qquad \bar{y}=\tau \omega =\sqrt{y^{2}-a^{2}}.
\end{equation*}

In fig. \ref{2peak} we compare Re$[g(x,y,a)]$ to Re$[g^{\prime }(x,y,a)]$
for $y=40$, $a=-1/10$, $-1$ and $-10$. We see that the approximation (\ref%
{gp}) captures the effects of the restriction to finite $\tau $ much better
when $|a|$ is large, while the approximation (\ref{repro}) tends to smear
them out. It is also easy to show that the real parts are not positive when $%
a$ is positive.

We can estimate the total effective width $\Gamma _{\text{tot}}$ of the
dressed propagator by means of twin BW approximations, obtained by replacing 
$f$ with the function $f_{\text{BW}}$ of (\ref{BWa}) inside (\ref{repro}) or
(\ref{gpp}). We assume that $y$ and $\bar{y}$ are large, to avoid
superpositions between the particle and antiparticle peaks. Making the
replacement $f\rightarrow $ $f_{\text{BW}}$ in $g(x,y,a)$, the shift $%
x\rightarrow x-ia$ in (\ref{effGamma}) gives 
\begin{equation}
\Gamma _{\text{tot}}\simeq \Gamma _{\text{eff}}-\frac{2a}{\tau }=\frac{16}{%
3\tau }-2\text{Im\hspace{0.01in}}[\tilde{\omega}_{\text{ph}}]\simeq \frac{16%
}{3\tau }+\frac{m_{\text{ph}}\Gamma }{\omega _{\text{ph}}}\simeq \frac{16}{%
3\tau }+\Gamma ,  \label{Gammatot}
\end{equation}%
the last-but-one approximation being for $\Gamma \ll m_{\text{ph}}$, and the
last one being at rest.

These results prove that the radiative corrections generate a shift into the
physical half plane. The effective width $\Gamma _{\text{eff}}$ of the free
propagator, due to the restriction to finite $\tau $, is enlarged to the
total width $\Gamma _{\text{tot}}$ of the dressed propagator by an amount
proportional to the usual width $\Gamma $ at $\tau =\infty $.

\section{Purely virtual particles}

\label{PV}\setcounter{equation}{0}

In this section we study purely virtual particles $\chi $, taking $\Delta
m^{2}=0$ again for simplicity. As recalled in the introduction, purely
virtual particles are introduced by removing the on-shell contributions of
ordinary particles, or ghosts, from the diagrams, perturbatively and to all
orders. If we do this on the Feynman propagator (\ref{pinf}), we lose $\pi
\delta (k^{2}-m^{2})$ and remain with 
\begin{equation*}
\mathcal{P}\frac{i}{k^{2}-m^{2}}=\frac{i}{2\omega }\left( \mathcal{P}\frac{1%
}{e-\omega }-\mathcal{P}\frac{1}{e+\omega }\right) .
\end{equation*}%
where $\mathcal{P}$ denotes the Cauchy principal value. The first pole
gives, after Fourier transform,%
\begin{equation*}
G_{\text{pv}}^{+}(t,t^{\prime })=\frac{\theta (t-t^{\prime })-\theta
(t^{\prime }-t)}{2}\frac{\mathrm{e}^{-i\omega (t-t^{\prime })}}{2\omega }.
\end{equation*}%
The second pole gives $G_{\text{pv}}^{-}(t,t^{\prime })=G_{\text{pv}%
}^{+}(t^{\prime },t)$.

The diagrams we are considering do not have $\chi $ legs inside loops (the
self-energy $\Sigma $ being treated as a whole), so the $\chi $ free
propagator is everything we need. Working out the Fourier transform $\tilde{G%
}_{\tau \hspace{0.01in}\text{pv}}^{+}(e,e^{\prime })$ of $G_{\tau \hspace{%
0.01in}\text{pv}}^{+}(t,t^{\prime })\equiv \Pi _{\tau }(t)G_{\text{pv}%
}^{+}(t,t^{\prime })\Pi _{\tau }(t^{\prime })$, defined as in formula (\ref%
{fu}), with $t_{\text{f}}=-t_{\text{i}}=\tau /2$, we find the result (\ref%
{Gtauplus}) with $f(z)$ replaced by%
\begin{equation*}
f_{\text{pv}}(z)=\frac{\sinh (z)-z}{z^{2}}.
\end{equation*}%
Hence, by (\ref{undrepro}), the total propagator reads%
\begin{equation*}
\tilde{G}_{\tau \hspace{0.01in}\text{pv}}(e,e^{\prime })\simeq \frac{2\tau
^{2}\sin \left( \frac{e+e^{\prime }}{2}\tau \right) }{e+e^{\prime }}h_{\text{%
pv}}(x_{+},x_{-}),\qquad h_{\text{pv}}(x_{+},x_{-})\equiv \frac{f_{\text{pv}%
}(ix_{+})+f_{\text{pv}}(-ix_{-})}{x_{-}-x_{+}}.
\end{equation*}%
The key function is now 
\begin{equation*}
f_{\text{pv}}(ix)=i\frac{x-\sin (x)}{x^{2}},
\end{equation*}%
which satisfies the important bound%
\begin{equation}
|f_{\text{pv}}(ix)|\leqslant \frac{1}{\pi }.  \label{bound}
\end{equation}

As in (\ref{gp}), the dressed propagator is a geometric series%
\begin{equation}
\frac{(e+e^{\prime })\tilde{G}_{\tau \text{d}}^{\prime \hspace{0.01in}\text{%
pv}}(e,e^{\prime })}{2\tau ^{2}\sin \left( \frac{e+e^{\prime }}{2}\tau
\right) }\simeq h_{\text{pv}}(x_{+},x_{-})\sum_{n=0}^{\infty }\left(
-m\Gamma \tau ^{2}h_{\text{pv}}(x_{+},x_{-})\right) ^{n}\equiv g_{\text{pv}%
}^{\prime }(x,y,a),  \label{b2}
\end{equation}%
but we cannot resum it without checking its actual convergence. The reason
is that the prescription for purely virtual particles is not analytic \cite%
{Piva,LWgrav}, so we cannot advocate analyticity to justify the continuation
of the sum from its convergence domain to the rest of the complex plane, as
we normally do for physical particles.

The bound (\ref{bound}) ensures that there is a situation where the series
is always convergent (on the real axis). It occurs when the quantity raised
to the power $n$ in the sum of (\ref{b2}) has a modulus that is always
smaller than 1. In turn, this requires $\gamma =\tau m\Gamma /(2\omega )<\pi
/2$, which is true for every energy $e$ and every frequency $\omega $, if $%
\tau \Gamma <\pi $. Thus, it is sufficient to assume 
\begin{equation}
\Delta E\equiv \frac{\pi }{2\tau }>\frac{\Gamma }{2},  \label{condo}
\end{equation}%
to obtain%
\begin{equation}
g_{\text{pv}}^{\prime }(x,y,a)=\frac{f_{\text{pv}}(i(x-\bar{y}))+f_{\text{pv}%
}(-i(x+\bar{y}))}{2\bar{y}\left( 1+\gamma f_{\text{pv}}(i(x-\bar{y}))+\gamma
f_{\text{pv}}(-i(x+\bar{y}))\right) }.  \label{proppv}
\end{equation}

When $\bar{y}$ is large, the \textquotedblleft particle\textquotedblright\
and \textquotedblleft antiparticle\textquotedblright\ contributions separate
well enough, and we can write%
\begin{equation*}
g_{\text{pv}}^{\prime }(x,y,a)\simeq \frac{\tilde{f}_{\text{pv}}(x-\bar{y}%
,\gamma )+\tilde{f}_{\text{pv}}(x+\bar{y},\gamma )}{2\bar{y}},\qquad \tilde{f%
}_{\text{pv}}(x,\gamma )=\frac{f_{\text{pv}}(ix)}{1+\gamma f_{\text{pv}}(ix)}%
.
\end{equation*}%
It is easy to prove that the twin peaks of Re$[\tilde{f}_{\text{pv}%
}(x,\gamma )]$ occur at $x=\pm \pi $, and have Re$[\tilde{f}_{\text{pv}}(\pm
\pi ,\gamma )]=\gamma /(\pi ^{2}+\gamma ^{2})$. Thus, $\gamma $ (hence $%
\Gamma $) is related to the heights of the peaks, while their positions are
universal. The stationary points of Re$[\tilde{f}_{\text{pv}}(x,\gamma )]$
are at $x=n\pi $, $n$ = odd, and have heights $\gamma /(n^{2}\pi ^{2}+\gamma
^{2})$.

\begin{figure}[t]
\begin{center}
\includegraphics[width=7truecm]{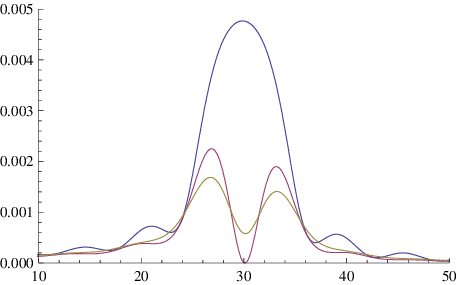}
\end{center}
\par
\vskip-.5truecm
\caption[Comparison between the dressed propagators of physical (in read)
and purely virtual (in blue) particles in a situation ($\protect\gamma =3/2$%
) where we can trust the resummation of the perturbative expansion for both]{%
Comparison between the dressed propagators Re$[g^{\prime }(x,y,a)]$ and Re$%
[g_{\text{pv}}^{\prime }(x,y,a)]$ of physical (in blue) and purely virtual
(in red and green) particles in a situation ($\protect\gamma =3/2$) where we
can trust the resummations for both. We have taken $\bar{y}=30$. The green
plot is the convolution (\protect\ref{convo})}
\label{Comparison}
\end{figure}

In fig. \ref{Comparison} we compare the properties of physical and purely
virtual particles through the functions Re$[g^{\prime }(x,y,a)]$ and 
\begin{equation}
\text{Re}[g_{\text{pv}}^{\prime }(x,y,a)]=g_{\text{pv}}^{\prime \ast
}(x,y,a)(m\tau ^{2}\Gamma )g_{\text{pv}}^{\prime }(x,y,a),  \label{pvdecay}
\end{equation}%
taking $\bar{y}=30$ and $\gamma =3/2$. The right-hand side of expression (%
\ref{pvdecay}) looks like the decay rate\ of the purely virtual particle $%
\chi $, because it is the product of the propagator, times (minus the real
part of) the bubble diagram, times the conjugate propagator (times a further
factor $\tau ^{2}$, introduced for convenience). Since $\chi $ does not
exist on the mass shell, the expression \textquotedblleft decay
rate\textquotedblright\ just refers to the existence of a channel mediated
by it.

Fig. \ref{Comparison} includes the plot of the convolution%
\begin{equation}
\int_{-\infty }^{+\infty }\mathrm{d}u\frac{4}{(4u)^{2}+\pi ^{2}}\text{Re}[g_{%
\text{pv}}^{\prime }(x-u,y,a)]  \label{convo}
\end{equation}%
with a Lorentzian function of width $\pi /2=\Delta x=\tau \Delta E$. The
convolution is useful if we want to interpret $\Delta E$ as the resolving
power of our instrumentation on the energy.

Since we are working in a finite interval of time $\tau $, we are implying
that the resolving power $\Delta t$ on time itself is better than that: $%
\Delta t<\tau $. Then, the energy-time uncertainty relation $\Delta t\Delta
E\gtrsim 1$ tells us that the uncertainty $\Delta E$ on the energy is bigger
than $\sim 1/\tau $. The best situation is when that uncertainty is close to
its minimum value, which is approximately equal to the $\Delta E$ defined in
formula (\ref{condo}). Thus, we can view the condition (\ref{condo}) as a
condition on the resolving power on the energy. If we resolve the energies
too well, we cannot have enough precision in time to claim that we are
working in a finite interval $\tau $. The convoluted profile of fig. \ref%
{Comparison} is probably closer to what we can see experimentally. 

The plots show rather different phenomenological behaviors: while physical
particles exhibit the usual peak, purely virtual particles show two smaller
humps. What is important is that the difference between the two cases is
experimentally testable, at least in principle.

Qualitatively, we may expect similar differences at $\tau =\infty $. However
we cannot make this statement rigorous, because when $\tau $ grows we
eventually violate (\ref{condo}) and enter a nonperturbative region, where
we cannot trust the resummation for purely virtual particles. At the
nonperturbative level, the condition (\ref{condo}) might turn into an
uncertainty relation of new type \cite{fSelfK}, a \textquotedblleft peak
uncertainty\textquotedblright\ $\Delta E>\Gamma /2$, telling us that, when
we approach the peak region $k^{2}\sim m^{2}$ of a purely virtual particle
too closely, identical experiments may give different results.

If purely virtual particles with relatively small masses exist in nature,
the predictions of this section could be tested exprimentally. Standard
model extensions that are worth of attention, in this context, have been
studied in refs. \cite{Tallinn1,Tallinn2}. Those models (which violate the
bound (\ref{condo}), because they have $\tau =\infty $) can be used without
modifications for qualitative tests. Consider, for example, processes that
involve exchanges of purely virtual particles, like $Z\rightarrow \mu \mu
\mu \mu $ \cite{Tallinn2}: once we reach enough precision, it should be easy
to realize that the shapes of the plots are more similar to the red and
green curves of fig. \ref{Comparison}, rather than the blue curve. For
quantitative tests, we need to extend the predictions of \cite%
{Tallinn1,Tallinn2} to a $\tau $ that is sufficiently small. The results of
this paper and \cite{MQ} give us the techniques we need, to achieve that
goal. 

More generally, the restriction to finite $\tau $, as well as the
restriction to a compact space manifold $\Omega $, can be used to amplify
effects that are otherwise too tiny to be observed, taking advantage of the
nontrivial interplay between the observed process and the external
environment (in particular, through the boundary of $\Omega $).

\section{The problem of the muon (unstable particles vs resonances)}

\label{muon}\setcounter{equation}{0}

In this section we study the problem of describing the muon decay in quantum
field theory. Since the muon is unstable, the right framework is not the one
at $\tau =\infty $, because a too large $\tau $ gives the muon enough time
to decay and, strictly speaking, makes it unobservable. If we ignore this
fact and insist on describing the muon decay at $\tau =\infty $, quantum
field theory retaliates by generating mathematical inconsistencies \cite%
{fSelfK}.

The point is that we are demanding something that violates the uncertainty
principle: as stressed before, if we want to resolve a finite time (the muon
lifetime in this case), we must have a finite time uncertainty $\Delta t$,
which needs a nontrivial uncertainty $\Delta E$ on the energy. There are no
such things in quantum field theory at $\tau =\infty $. On the other hand, a
finite $\tau $ implies a finite time uncertainty $\Delta t<\tau $, so
quantum field theory on a finite time interval is better equipped to address
the problem we are considering. Moreover, if we want to be able to see the
muon, we must have $\tau <\omega \tau _{\mu }/m_{\mu }$, where $\tau _{\mu
}=1/\Gamma _{\mu }$ is the muon lifetime at rest, $m_{\mu }$ is the muon
mass and $\omega /m_{\mu }$ is the boost factor, which is crucial to make
the muon live longer.

Let us imagine a process where certain incoming particles $X$ collide and
produce the unstable particle, or resonance, we want to study, which we
denote by $\phi $. The total cross section $\sigma _{\text{tot}}$ can be
split into the sum of the cross section for the production of $\phi $ itself
(in which case $\phi $ does not decay during the process, and is the sole
outgoing state), and the cross section for the products of the $\phi $
decay. The optical theorem tells us that $\sigma _{\text{tot}}$ is
proportional to the real part of the forward scattering amplitude $%
X\rightarrow \phi \rightarrow X$, which receives its most important
contribution around the $\phi $ peak from the $\phi $ dressed propagator.
For example, we can take $X=e^{+}e^{-}$ and $\phi =Z$ to describe the $Z$
production at LEP.

The $\phi $ propagator we work with is the function%
\begin{equation}
g^{\prime }(x,y,a)=\frac{h(x_{+},x_{-})}{1+2\gamma \bar{y}h(x_{+},x_{-})},
\label{gppp}
\end{equation}%
from formula (\ref{gpp}). Its real part can be written as the sum Re$%
[g^{\prime }]=(\Omega _{\text{particle}}+\Omega _{\text{decay}})/\tau ^{2}$
of the two terms 
\begin{equation}
\frac{\Omega _{\text{particle}}}{\tau ^{2}}=\frac{h_{1}}{|1+2\gamma \bar{y}%
h|^{2}},\qquad \qquad \frac{\Omega _{\text{decay}}}{\tau ^{2}}=\frac{2\gamma 
\bar{y}|h|^{2}}{|1+2\gamma \bar{y}h|^{2}}=g^{\prime \hspace{0.01in}\ast
}(m\tau ^{2}\Gamma )g^{\prime },  \label{Omega}
\end{equation}%
where $h$ stands for $h(x_{+},x_{-})$ and $h_{1}$ is its real part. The
factor $1/\tau ^{2}$ takes care of the analogous factor appearing on the
left-hand side of (\ref{gp}). The reason behind the separation (\ref{Omega})
is relatively simple to understand: $\Omega _{\text{decay}}$, which captures
the $\phi $ decay, is the part proportional to the self-energy itself (i.e.,
proportional to $\Gamma $, in our approximation), while $\Omega _{\text{%
particle}}$, which captures the particle observation, is the rest. The
detailed resummation of the diagrams involved in the two cases can be found
in \cite{fSelfK}.

The structure of $\Omega _{\text{decay}}/\tau ^{2}$ matches the one of (\ref%
{pvdecay}), while the first expression has no analogue in the case of purely
virtual particles (which admit no particle observation, by definition, but
just a \textquotedblleft decay\textquotedblright\ channel).

We want to study $\Omega _{\text{particle}}$ and $\Omega _{\text{decay}}$ in
two limiting situations of physical interest: unstable particles and
resonances.

Although we just need to take $\tau $ smaller than the boosted muon lifetime 
$\omega \tau _{\mu }/m_{\mu }$, it is convenient to take $\tau \ll \omega
\tau _{\mu }/m_{\mu }$, both because it is realistic to do so, but also
because it simplifies the results. Furthermore, in all the colliders built,
or planned, so far, the muon mass $m_{\mu }$ is much larger than the
resolving power on the energy, so we may assume 
\begin{equation}
m_{\mu }\gg \frac{1}{\tau }\gg \frac{m_{\mu }}{\omega \tau _{\mu }}=\frac{%
m_{\mu }\Gamma _{\mu }}{\omega }.  \label{approx}
\end{equation}%
It is easy to prove the inequality $|h(x_{+},x_{-})|\leqslant 1/(2\bar{y})$,
which implies that under the assumptions (\ref{approx}), the denominator of $%
g^{\prime }(x,y,a)$ in (\ref{gppp}) can be approximated to one, and the
function $g^{\prime }(x,y,a)$ can be approximated to its free value $%
h(x_{+},x_{-})$ (which implies $\Omega _{\text{decay}}\rightarrow 0$).

Furthermore, the conditions (\ref{approx}) imply $\bar{y}\gg 1$. Then it is
easy to prove\footnote{%
We need to take $|\tau e|$ large and comparable to $\bar{y}=\tau \omega $,
otherwise we miss the delta function support. Basically, we are rescaling $e$
and $\omega $ by a common factor, and letting it tend to infinity. At the
same time, we keep $\tau $ fixed.}, from the second limit of (\ref{distro}),
that 
\begin{equation}
\Omega _{\text{particle}}\simeq \pi \delta (e^{2}-\omega ^{2}).
\label{correct}
\end{equation}%
In other words, $\Omega _{\text{particle}}$ tends to the delta function that
describes the muon observation, while $\Omega _{\text{decay}}$ tends to zero.

Note that we have not taken $\tau $ to infinity to prove this result.
Actually, it is impossible to obtain it by working at $\tau =\infty $ \cite%
{fSelfK}, because in that case 
\begin{equation*}
\tau ^{2}g^{\prime }(x,y,a)\rightarrow \frac{i}{e^{2}-\omega ^{2}+i(\epsilon
+m_{\mu }\Gamma _{\mu })},
\end{equation*}%
which means%
\begin{equation}
\Omega _{\text{particle}}\rightarrow \frac{\epsilon }{(e^{2}-\omega
^{2})^{2}+(\epsilon +m_{\mu }\Gamma _{\mu })^{2}},\qquad \qquad \Omega _{%
\text{decay}}\rightarrow \frac{m_{\mu }\Gamma _{\mu }}{(e^{2}-\omega
^{2})^{2}+(\epsilon +m_{\mu }\Gamma _{\mu })^{2}}.  \label{formu}
\end{equation}%
Since $\Gamma _{\mu }$ is nonzero and $\epsilon $ is a mathematical
artifact, we get $\Omega _{\text{particle}}\rightarrow 0$, while $\Omega _{%
\text{decay}}$ tends to the Breit-Wigner function of a resonance. Normally,
people confuse $\Omega _{\text{particle}}$ and $\Omega _{\text{decay}}$, and
say that, because the muon width $\Gamma _{\mu }$ is very small, one can let
it tend to zero in $\left. \Omega _{\text{decay}}\right\vert _{\epsilon =0}$%
, which gives $\pi \delta (e^{2}-\omega ^{2})$. However, the desired delta
function should not come from $\Omega _{\text{decay}}$ (it would be like
resuscitating the muon by making it eternal after its decay): it must come
from $\Omega _{\text{particle}}$. This can happen only at $\tau <\infty $,
as in (\ref{correct}).

In the case of a resonance, like the $Z$ boson, there is no reason why we
should keep $\tau $ finite, since in all the experiments of collider
physics, so far, the $Z$ lifetime $\tau _{Z}$ is much shorter than the
interval $\tau $ separating the incoming particles from the outgoing ones
(we are very far from observing the $Z$ boson directly): 
\begin{equation}
m_{Z}>\Gamma _{Z}=\frac{1}{\tau _{Z}}\gg \frac{1}{\tau }.  \label{approZ}
\end{equation}

This means that we can use the formulas (\ref{formu}) with $\mu \rightarrow
Z $, where $\Omega _{\text{particle}}$ correctly gives zero, while $\Omega _{%
\text{decay}}$ tends to the right Breit-Wigner formula.

The processes observed in colliders fall in one of the situations just
described, where the particle lifetimes $\tau _{\phi }$ are much longer, or
much shorter than $\tau $. If we want to test formulas such as (\ref{gppp})
beyond the approximations considered above, $\tau $ must be comparable with $%
\tau _{\phi }$, and the energy precisions must be comparable with the
widths. We can reach the required $\tau $ with muons and tauons (a tauon
with an energy equal to the maximum LHC energy, 13.6TeV, travels 66
centimeters). It is much harder to reach the required energy resolutions,
because a huge gap separates the widths of the known renonances from the
ones of the long-lived unstable particles: there are 19 and 12 orders of
magnitude between the width of the $Z$ boson and the ones of the muon and
tauon, respectively. The conclusion is that, right now, it is hard to figure
out realistic intermediate situations between the two limits that we have
consided. Still, it is worth to point out that, if a chance of that type
ever becomes available, a way to test formulas like (\ref{gppp}) is to count
only particle traces with specific features, e.g., longer/shorter than some
given length $\ell $ (the critical value being $\ell \sim \tau _{\phi }\bar{E%
}/m_{\phi }$, where $\bar{E}$ is the mean particle energy and $m_{\phi }$ is
its mass). Plotting the data as functions of the muon energy, one should
find a distribution with a width that is larger than $\Gamma $, as predicted
by (\ref{Gammatot}).

\section{Conclusions}

\label{conclusions}\setcounter{equation}{0}

We have studied the propagators of physical and purely virtual particles in
quantum field theory in a finite interval of time $\tau $, and on a compact
manifold $\Omega $. In the free-field limit, the typical pole $1/z$ is
replaced by the entire function $f(z)=(e^{z}-1-z)/z^{2}$. The shape of the
latter on the real axis $z=ix$ reminds the one of a Breit-Wigner function,
with an effective width equal to $16/(3\tau )$. The two functions are very
different in the rest of the complex plane.

When we include the radiative corrections, the key function remains $f(z)$,
but it is shifted into the physical half plane. The width is enlarged by an
amount equal to $\Gamma $ (the usual width at $\tau =\infty $). The real
part of the propagator is always positive, in agreement with unitarity.

We have studied the case of purely virtual particles, and showed that, for $%
\tau $ small enough ($\tau <\pi /\Gamma $), there is an arrangement where
the geometric series of the self-energies is always convergent. The key
reason is that the function $f(z)$ is bounded on the real axis and on the
physical half plane. In that situation, it is possible to rigorously resum
the series into the dressed propagator, and compare the result with what we
find in the case of physical particles. The plots differ in ways that can in
principle be tested: physical particles are characterized by the usual,
single peak; instead, purely virtual particles are characterized by two twin
peaks, which are separated from one another in a universal way, and have
heights that depend on the width of the particle.

Finally, we have investigated the effects of the restriction to finite $\tau 
$ on the problem \textquotedblleft muon vs $Z$ boson\textquotedblright\
(i.e., unstable particles vs resonances). It is crucial to work at $\tau
<\infty $, if we want to properly explain the observation of an unstable
particle. Once we do that, the muon observation emerges naturally from the
right physical process. In particular, there is no need to confuse the
observation of a particle with the observation of its decay products, and
pretend that the particle resuscitates after its decay (which is basically
how one normally adjusts the matter by sticking to $\tau =\infty $). The
results confirm those argued in ref. \cite{fSelfK} on general grounds.

Examples of time-dependent problems where it might be interesting to use the
techniques studied here and in \cite{MQ} are neutrino oscillations and kaon
oscillations, as well as phenomena of the early universe and quark-gluon
plasma. Hopefully, the investigation carried out here can stimulate the
search for ways to overcome the paradigms that have dominated the scene in
quantum field theory since its birth, by searching for purely virtual
particles, on one side, and outdoing the $S$ matrix and the diagrammatics
based on time ordering, on the other side. In this spirit, it may be
interesting to merge the results with those of approaches like the
Schwinger-Keldysh \textquotedblleft in-in\textquotedblright\ formulation,
which applies to initial value problems, and also involves a diagrammatics
that is different from the standard \textquotedblleft
in-out\textquotedblright\ one.

\end{document}